\documentclass{aa}
\usepackage{graphicx}
\usepackage{supertabular}
\usepackage{txfonts}
\usepackage{natbib}

\bibpunct{(}{)}{;}{a}{}{,}

\begin{document} 

\title{SDSS J1553+0056: A BALQSO mimicking a Lyman-break galaxy 
\thanks{Based on
observations obtained with {\sc Uves} at the ESO Very Large Telescope,
Paranal, Chile (proposal No.~073.A-0386(A)).}}

\author{I. Appenzeller\inst{1} 
      \and O. Stahl\inst{1} 
      \and C. Tapken\inst{1}
      \and D. Mehlert\inst{1}
      \and S. Noll\inst{2}} 

\offprints{I. Appenzeller}
\institute{Landessternwarte, K\"onigstuhl, D 69117 Heidelberg, Germany
\and Max-Planck-Institut f\"ur extraterrestrische Physik, Giessenbachstr.,
D 85741 Garching, Germany
}
\date{Received 13 December 2004/ Accepted 12 January 2005}

\abstract{Using the UVES echelle spectrograph at the ESO VLT we obtained
  high-resolution (R = 40\,000) spectra of the object SDSS J1553+0056, which
  has been identified in the literature alternatively as a high-redshift
  quasar or as a Lyman-break galaxy (LBG). Although low-resolution spectra of
  SDSS J1553+0056 closely resemble  those of LBGs, our high-resolution
  spectra allow us to identify this object unambiguously as a LoBAL quasar,
  probably belonging to the rare FeLoBALQSO class.  Based on our spectrum we
  discuss how misidentifications of such objects from low-resolution spectra
  can be avoided.
        
} 
\authorrunning{I. Appenzeller et al.}  
\titlerunning{SDSS J1553+0056}
\maketitle

\section{Introduction}\label{introduction}
In addition to providing a large sample of galaxy spectra, the Sloan Digital
Sky Survey \citep{2000AJ....120.1579Y} has up to now resulted in  more than 
50 000 low-resolution spectra of quasars.  These spectra have been published in
various SDSS data releases
\citep{2002AJ....123..485S,2003AJ....126.2081A,2004AJ....128..502A} and
cataloged by \citet{2002AJ....123..567S,2003AJ....126.2579S}. As pointed out
by \citet{2004AJ....127..576B} and \citet{2004ApJ...600L..19B}, the
low-resolution spectra of some of the objects classified as quasars by the
SDSS pipeline closely resemble spectra normally observed for high-redshift
starburst galaxies.  Therefore, \citet{2004ApJ...600L..19B} suggest that
these objects are either luminous starburst galaxies or very peculiar
BALQSOs and argue that
some of the high-redshift objects in the SDSS quasar catalog are actually
high-redshift galaxies or ``Lyman-break galaxies'' (LBGs). 

If this
suggestion is correct, the objects listed by \citet{2004ApJ...600L..19B} would
be among the brightest high-redshift starburst
galaxies known, both apparently and absolutely. In fact these objects 
are bright enough to be observed with
high spectral resolution using efficient Echelle spectrographs at 8-10 m
telescopes. With the single exception of the gravitationally lensed galaxy MS
1512-cB58
\citep{1996AJ....111.1783Y,2000ApJ...528...96P,2002ApJ...569..742P,2002ApJ...567..702S}
no high-resolution spectra of Lyman-break galaxies have been published so far,
and almost all our knowledge about the structure, physical state, and
composition of these objects has been inferred from low-resolution spectra.
Therefore, we obtained high-resolution echelle spectra of the $z=2.63$ object
SDSS J155359.96+005641.4 from the list of \citet{2004ApJ...600L..19B}. 
The aim of our
observations was to clarify the physical nature of SDSS J1553+0056 and to gain
information on the physics of such luminous objects showing (at low spectral
resolution) Lyman-break-galaxy spectra.

\section{Observations, data reduction, and basic spectroscopic results
}\label{observations}

The observations were carried out in service observing mode between April and
June 2004 with {\sc Uves}, the {\bf U}ltraviolet and {\bf V}isual {\bf
  E}chelle {\bf S}pectrograph at the Nasmyth platform B of ESO's VLT UT2
(Kueyen) on Cerro Paranal, Chile. For all spectra the red channel and the CD
No. 3 were used. In total 8 individual spectra covering the spectral range
4160 \AA\ -- 6200 \AA\ and 5 spectra covering the overlapping range 4800\AA\ 
-- 6820 \AA\ were obtained. Thus, in the rest frame of SDSS J1553+0056 our
combined spectra cover the wavelength range 1145 -- 1880 \AA.  The projected
slit width was 1 arc sec, resulting in a spectral resolution of 
$\lambda /\Delta \lambda  \approx $ 40\,000 which
corresponds to a velocity resolution of 7.5 km s$^{-1}$. The exposure times
of the individual frames varied between 28 and 50 minutes, the total
integration time for the spectral overlap region amounted to about 10.5 hours.
Three of the individual exposures with a total exposure time of more than
2 hours were, however, obtained under poor observing conditions 
(outside specifications)
and, therefore, contain little information.  The average FWHM seeing for the
observations was about 0.8 arcsec, and extrema were 0.5 and 1.8 arcsec.

\begin{figure}
\resizebox{\hsize}{!}{\includegraphics[angle=0]{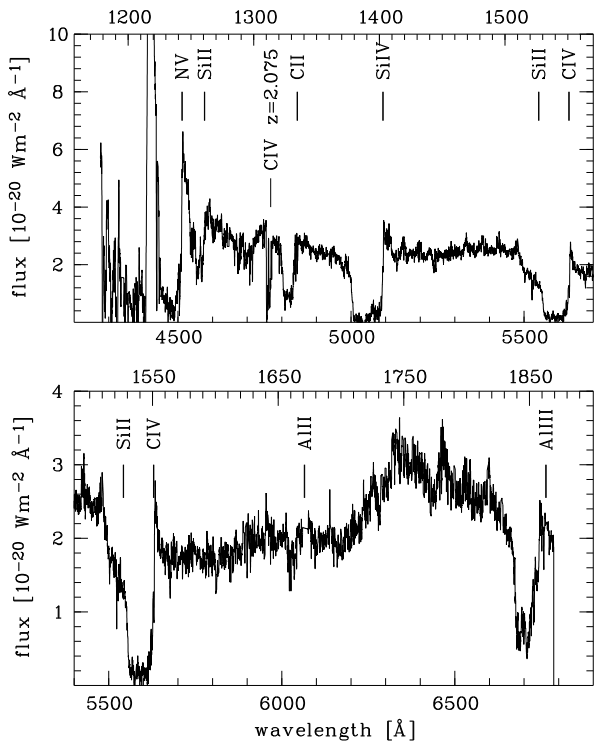}}
\caption[]{The complete spectrum of SDSS J1553+0056. The top axis plots
rest-frame  wavelengths assuming a systemic redshift of $z$ = 2.6327. The
bottom axis gives the observed wavelengths.}
\label{completespectrum}
\end{figure}

Because of the low signal in parts of our spectra the standard ESO pipeline
software for UVES turned out to be unsuitable for our data; therefore, the
reduction was carried out using special software initially developed by us for
the reduction of FEROS echelle spectra \citep{1999oisc.conf..331S}. In a first
step we carried out a two-dimensional reduction of the spectral frames. Then
the echelle orders were merged and one-dimensional spectra extracted. For
this procedure all spectral regions with defects or too low an S/N were
omitted. The wavelength scale was then converted to heliocentric vacuum
wavelengths, and spectra from the individual exposures were co-added with
equal weight. Outside the absorption troughs the resulting merged spectrum has
a continuum signal-to-noise ratio (per spectral resolution element) of about
10.

\begin{table}
\caption{Full widths (at continuum level) of the absorption troughs of 
resonance lines. For the doublets 
the measured width was reduced by the doublet separation to get values
corresponding approximately to the individual doublet components.}
\label{Linewidths}
\begin{center}
\small{
\begin{tabular}{llll}
\hline
Ion & $\lambda _{lab,vac}$ (\AA ) &  trough width (km s$^{-1}$) & remark\\
\hline
N\,{\sc v} & 1238.82,1242.80 & 4200 ? & bl. Ly$\alpha $ \\
Si\,{\sc ii} & 1260.42 & 2994 & \\
C\,{\sc ii} & 1334.53 & 2413 & \\
Si\,{\sc iv} & 1393.76,1402.77 & 3997 & \\
C\,{\sc iv} & 1548.20,1550.77 & 4624 & bl. Si\,{\sc ii} (UV2)\\
Al\,{\sc ii} & 1670.79 & 2471 & bl. Fe\,{\sc ii} (UV40 -- 42)\\
Al\,{\sc iii}  & 1854.72,1862.79 & 2158 &  \\
\hline
\end{tabular}
}
\end{center}
\end{table}

Since the spectra were taken under variable observing conditions, no direct
flux calibration of our UVES spectra was attempted. Instead we used the
archive SDSS spectrum of SDSS J1553+0056 to correct for the UVES instrumental
response and to convert the spectrum to flux units.  The corresponding total
flux-calibrated spectrum is reproduced in Fig.~\ref{completespectrum}. For
plotting this figure the effective spectral resolution has been reduced to R
$\approx$ 5000 using a median filter.

\section{The physical nature of SDSS J1553+0056}\label{nature}

\begin{figure}
\resizebox{\hsize}{!}{\includegraphics[angle=0]{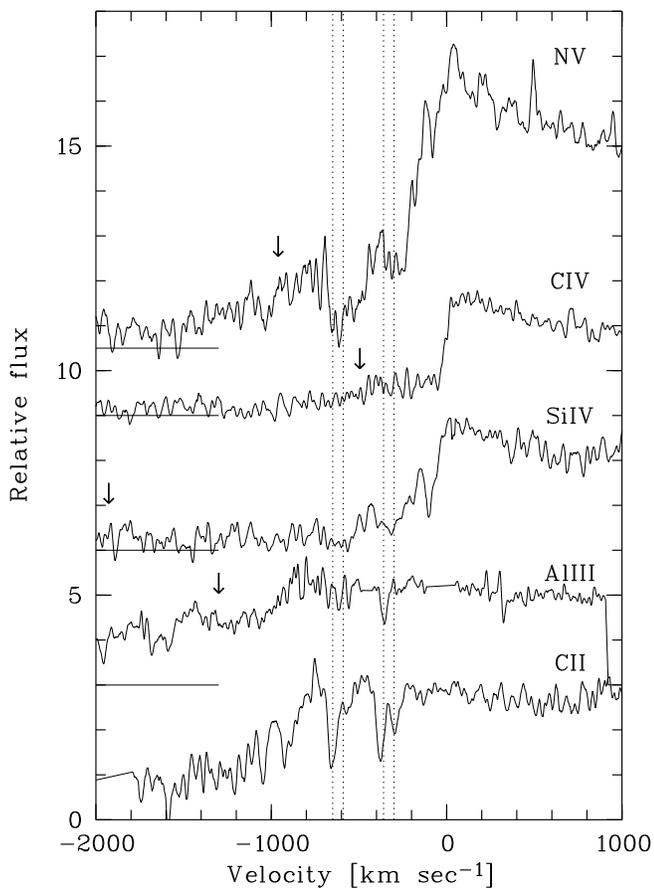}}
\caption[]{Profiles of the red edges of several resonance-line absorption 
troughs. The velocity scale assumes a systemic redshift of 2.6327. In the case 
of doublets the velocity refers to the red doublet component. The 
($v = 0$) positions of the blue doublet components are indicated by
vertical arrows. In the case 
of C\,{\sc ii} the velocity scale 
corresponds to the mean wavelength of the two fine structure components
C\,{\sc ii}$^{*}$ $\lambda \lambda $ 1335.66, 1335.71 of C\,{\sc ii} (UV1). 
Features caused by the ground state $\lambda $ 
1334.53 line of this multiplet are shifted (by about 260 km s$^{-1}$)
to apparently more negative velocities. As a result the 
$-360$ km s$^{-1}$ component of the  1334.53 \AA\ line overlaps with
the $-650$ km s$^{-1}$ component of C\,{\sc ii}$^{*}$, while the $-650$ km s$^{-1}$ 
feature of the 1334.53 line occurs at the apparent velocity
$-910$ km s$^{-1}$. The apparent common absorption features mentioned in 
the text are indicated by dotted vertical lines.}
\label{resonancelines}
\end{figure}

As shown by Fig.~\ref{completespectrum} the spectrum of SDSS J1553+0056 is
dominated by broad absorption troughs of resonance lines and blends. The
shapes (see Fig.~\ref{completespectrum}) and widths (see Table~1) of these
lines are clearly reminiscent of the spectra of the Broad Absorption Line
(BAL) quasars \citep[see
e.g.][]{1991ApJ...373...23W,2002ApJS..141..267H,2003AJ....126...53B,2003AJ....125.1711R}.
But apart from Ly$\alpha $ the spectrum of SDSS J1553+0056 shows no strong
emission lines.  Absorptions troughs corresponding to low-ionization resonance
lines are prominent. Thus, SDSS J1553+0056 obviously belongs to the LoBAL
sub-class of the BALQSOs as defined by \citet{1993ApJ...413...95V}. The red
edges (i.e. wavelengths corresponding to the onset of the sharp drop of
the flux in the red wings of the troughs) of the Ly$\alpha $, N\,{\sc v}
$\lambda$ 1242.80, Si\,{\sc iv} 1402.77, and C\,{\sc iv} 1550.77 absorption
troughs correspond to a common redshift of 2.6327 $\pm $ 0.0002. In the
following we shall assume, therefore, that this value  
approximates the
systemic redshift of this BALQSO.  All velocities listed below are given
in relation to this redshift.  As shown in Fig.~\ref{resonancelines}, the red
edges of the low-ionization troughs have smaller redshifts. But all
low-ionization lines show narrow (FWHM $ \leq $ 40 km s$^{-1}$) line
components with redshifts between the redshift of the high-ionization and the
low-ionization red edges.

In the case of resonance doublets, the edge profiles plotted in
Fig.~\ref{resonancelines} correspond to the red doublet components. For
Si\,{\sc iv} there is no significant overlap
by the corresponding blue doublet component in the velocity range 
given Fig. 2. In the case of Al\,{\sc iii},
N\,{\sc v}, and C\,{\sc iv} the doublet components overlap for $v \leq -1300,$
$\leq -960,$ and $\leq -500$ km s$^{-1}$, respectively. Because of these
overlaps and the low flux inside the troughs, the present discussion is
restricted to the red edges.

Some of the narrow low-ionization lines mentioned above are visible in
Fig.~\ref{resonancelines}. Since part of these narrow lines in
Fig.~\ref{resonancelines} are observed with the same profiles in different
echelle orders these features definitely are not artifacts. In the case of the
Al\,{\sc iii} resonance doublet, the large doublet separation 
and the relatively
high residual flux in the absorption trough (see Fig.~\ref{completespectrum})
allowed us to compare the edge profiles and the pattern of sharp absorption
lines of the two doublet components. As expected, the edge profiles and the
sharp line profiles of the two doublet components turned out to be identical.
On the other hand Fig. 2 and Table 1 show that for
lines not belonging to the same ion the edge profiles the sharp components of
the low-ionization lines and the widths of the trough profiles are clearly
different. There are some common properties, however, most profiles show
absorption features at $-300$ to $-360$ km s$^{-1}$ and $-590$ to 
$-650$ km s$^{-1}$. These features are
narrow in the low-ionization lines and broader for the high-ionization
species. The mean velocities of these velocity components, as
derived from averaging the velocities measured for the individual lines,
are indicated in Fig. 2 by dotted vertical lines. All troughs with 
sufficient rest flux show a 'window' of less
deep trough absorption near $-760$ km s$^{-1}$.

It cannot be excluded that the observed narrow absorption lines
described above are caused by foreground metal absorption systems with slightly
smaller redshifts.  However, the absence of corresponding narrow features at
the high-ionization lines, which only show broader features, 
seems to argue against this assumption.  Therefore,
we suggest that the sharp lines form in (still) relatively cool and dense
interstellar clouds of the ISM of the QSO host galaxy which so far have been
shielded from the central source and have not yet been significantly
accelerated by radiation from the central source. If this assumption is
correct, the observed structure of the red edges of resonance lines and
the presence of narrow line components with common velocities slightly
blueward of the high-ionization red edge velocity both seem to support 
suggestions
that BALQSO troughs result from the superposition of individual absorption
lines originating in a modest number of absorbing gas clouds that are
accelerated (and eventually dispersed) by a recently switched on AGN \citep[see
e.g.][]{1999MNRAS.310..913W,2002ApJS..141..267H}.  The observed profile
differences correlated with the ionization stage indicate that the absorbing
clouds at this evolutionary stage differ greatly in their physical parameters
and in their internal radiation and velocity fields.
 
\begin{figure}
\resizebox{\hsize}{!}{\includegraphics[angle=-90]{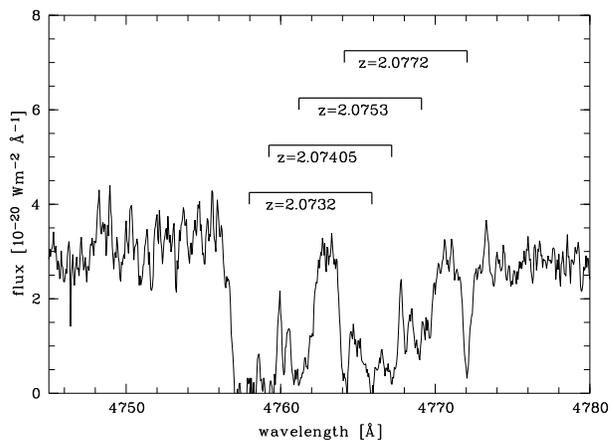}}
\caption[]{Profile of the complex 2.0729 $ < z < $ 2.0772 foreground 
C\,{\sc iv} doublet absorption system.}
\label{civlines}
\end{figure}

In addition to the narrow resonance absorption lines with redshifts close to
the assumed systemic velocity, at least three C\,{\sc iv} absorption systems
with smaller redshifts (at $z$ = 1.8679, 2.0413, and 2.0729 -- 2.0772) are
present. The complex profile of the C\,{\sc iv} doublet of the 2.0729 --
2.0772 system is reproduced in Fig.~\ref{civlines}. The Al\,{\sc ii} and
Al\,{\sc iii} resonance absorption lines of this system are also present in
our spectra.

Apart from the fine structure lines of C\,{\sc ii} (UV1) mentioned in Fig. 2, 
no other narrow absorption lines originating from excited atomic energy levels
could be detected. Moreover, a thorough search for lines typical of the
photospheric spectra of hot stars (such as O\,{\sc iv} 1343.35, C\,{\sc iii}
1427.85, and S\,{\sc v} 1501.76 \AA\ at $z \approx 2.633$) was not
successful, so that integrated stellar light from the QSO's host galaxy cannot
contribute much to the observed flux. The depth of the C\,{\sc iv} and
Si\,{\sc iv} trough shows that coverage of the continuum source by the
absorbing clouds must be close to complete.

As shown by Fig.~\ref{completespectrum} the red part of our spectrum contains
a region of high flux with some apparent peaks in the wavelength range 6300
-- 6600 \AA\ (corresponding to $\approx $ 1740 -- 1800 \AA\ in the QSO's
restframe) and a region of low flux between 5600 and 6300 \AA\ (1550 -- 1740
\AA, restframe).  These SED features are qualitatively similar to those
observed in certain FeLoBALQSOs, such as SDSS 0318-0600
\citep{2002ApJS..141..267H}, PSS J0052+2405 \citep{2003AJ....126...53B}, in
the composite FeLoBAL spectrum presented by
\citet[][ Fig.~1]{2003AJ....125.1711R}, and in the BALQSO
FDF-6233 \citep{2004A&A...418..885N}. As pointed out by
\citet{2002ApJS..141..267H} these features can be explained by the presence of
of Ni\,{\sc ii} and Fe\,{\sc ii} lines originating in part from excited atomic
energy levels. A detailed comparison with the UV multiplet tables shows that
modulation of the pseudo-continuum of SDSS J1553+0056 in the 
region 5600 -- 6300
\AA\ is qualitatively consistent with a superposition of merging 
broad absorption lines
originating from Si\,{\sc ii} (UV1), Ni\,{\sc ii} (UV2 to UV5), and Fe\,{\sc
  ii} (UV multiplets 8, 9, and 38 -- 46).  At least part of the apparent steps
in the flux distribution seem to coincide with the red limits of some of these
multiplets. But no narrow absorption components originating from excited
Fe\,{\sc ii} energy levels could be detected so the individual lines
appear to produce weak absorption troughs of unknown width.

The possible presence of broad Fe\,{\sc ii} absorption 
from excited energy levels
suggest that SDSS J1553+0056 is a member of the rare FeLoBALQSO class
\citep{1987ApJ...323..263H,1997ApJ...479L..93B,2000ApJ...538...72B,2001ApJ...561..645M}.
But our UVES spectra do not contain those lines 
of the Fe\,{\sc ii} multiplets UV1
and UV2 that define this class. The low-resolution spectrum of SDSS
J1553+0056 in the SDSS archives includes the Fe\,{\sc ii} (UV2) multiplet.  On
this spectrum an Fe\,{\sc ii} (UV2) absorption trough appears to be present,
but the corresponding spectral region coincides with telluric OH emission.
Hence, a definite confirmation of the FeLoBALQSO character of SDSS J1553+0056
will require higher resolution spectra in the red ($\lambda >$ 7000 \AA)
spectral range.

Compared to other FeLoBALQSOs, our target shows little evidence for dust
reddening.  The overall SED in the observed wavelength range is bluer than the
dereddened SED of the FeLoBALQSO SDSS 0318-0600 \citep{2002ApJS..141..267H}.


\section{Comparison with galaxy spectra and implications for LBG 
  searches}\label{implications} \citet{2004ApJ...600L..19B} suggested that 
SDSS J1553+0056 is a Lyman Break Galaxy on the basis of the 
high-ionization resonance absorption lines observed (thought to be stellar wind
features) and the apparent presence of narrow (ISM) absorption lines of
Si\,{\sc ii}, O\,{\sc i}, and C\,{\sc ii}, which are prominent in LBG
spectra. Moreover, for a QSO the SED of SDSS J1553+0056 is unusually 
flat and
resembles that of a slightly reddened LBG, and the Ly$\alpha $ line of SDSS
J1553+0056 appears rather narrow for a QSO.

Our high resolution spectra show that the narrow appearance of the Ly$\alpha $
line is due to absorption by the Ly$\alpha $ trough in the blue wing and by
the N\,{\sc v} trough in the red wing. The other emission lines appear
intrinsically weak in SDSS J1553+0056 and are partly absorbed by their
own troughs. These troughs show no or little detachment from 
the emission features.  Moreover,
low-ionization line emission is weakened on the low-resolution spectrum by
unresolved narrow absorption features. That the BAL absorption troughs
have been misidentified as stellar wind features is obviously due to their
modest width, which is near the lower limit  found for BALQSOs. However,
the observed trough widths are still inside the range of
2000 -- 20\,000 km s$^{-1}$ typical for this class of objects.
The particularly low
widths of the C\,{\sc ii} and Si\,{\sc ii} line troughs resulted in their
misidentification as ISM features. And the foreground $z \approx $ 2.07
C\,{\sc iv} system appearing by chance coincidence at the corresponding
wavelength mimicked the presence of a sharp ISM O\,{\sc i} absorption
feature.

LoBALQSOs, in particular FeLoBALQSOs, are relatively rare objects
\citep[see e.g.][]{2001ApJ...561..645M}. However the surface density on the
sky of high-redshift LBGs is also small, so at least at the high absolute
luminosity end, LBG identifications from low resolution spectra can be affected
by objects similar to SDSS J1553+0056. As pointed out by
\citet{2004ApJ...600L..19B} identifications of high-luminosity LBGs determine
the galaxy luminosity functions at high redshift, with significant
cosmological implications.  Thus, it is important to avoid such
misidentification.

On the basis of our results the most reliable separation between objects such
as SDSS J1553+0056 and LBGs from low resolution spectra appears to be the
continuum features caused by excited Fe\,{\sc ii} levels in the restframe 1550
-- 1800 \AA\ range, as such features are never observed in LBGs. Unfortunately
evaluation of these continuum features, while possible at very low spectral
resolution, requires very accurate flux calibration of the spectra, which is
difficult for faint objects. Moreover, in QSOs this feature seems to be
restricted to the rare FeLoBALQSOs.
   
If the high-ionization absorption troughs of a BALQSO are as narrow as or
narrower than in the case of SDSS J1553+0056 at low spectral resolution
misidentification with stellar wind lines which in certain types of hot stars
can reach similar widths probably cannot be excluded.  However, comparison
of our UVES spectrum with LBG spectra indicates that the strength of the
C\,{\sc ii} (UV1) absorption in the low-resolution spectrum of SDSS J1553+0056
(EW$_\mathrm{rest} >6$ \AA) could have been used to determine the nature of this
object. In LBGs this absorption is normally caused by ISM absorption. Such ISM
lines are usually narrow, which results in a low equivalent width ($\leq 4$
\AA\ in LBGs), independent of the resolution.
Thus, the high equivalent widths of low-ionization resonance lines, such as
the strong C\,{\sc ii} (UV1) absorption observed in the low-resolution
spectrum of SDSS J1553+0056 are telltale signs of the presence of 
the absorption troughs characteristic of LoBALQSOs.

Finally, we note that in composite spectra of representative samples of LBGs
(see e.g. \citealt{2003ApJ...588...65S} and \citealt{2004A&A...418..885N}),
the N\,{\sc v} stellar wind absorption feature is never as deep as observed in
SDSS J1553+0056. Therefore, all LBG spectra showing such strong 
blueshifted N\,{\sc v} absorption should be checked for the presence of 
other BALQSO indicators.

Application of the above criteria indicates that in addition to SDSS
J1553+0056 at least three more of the 6 objects in the 
\citet{2004ApJ...600L..19B} list could possibly be BALQSOs. Thus, it appears
desirable to clarify the nature of these objects by means of higher resolution
spectroscopy before conclusions are drawn from their tentative
identifications as LBGs.

\section{Conclusions}\label{discussion}

Our high-resolution spectroscopic observations of SDSS J1553+0056 have shown
that this object is not an LBG but instead a LoBALQSO that probably belongs to
the FeLoBALQSO class. Compared to other BALQSOs it has weak emission lines,
relatively narrow absorption troughs; and it shows (relative to other
FeLoBALQSOs) little evidence of dust reddening. Because of these properties
the low-resolution spectrum of SDSS J1553+0056 resembles those of
high-redshift starburst galaxies. However, there is no evidence for any
stellar light contribution to the observed flux from SDSS J1553+0056. 
From a comparison of our spectra with mean LBG spectra we conclude that on
low-resolution spectra objects similar to SDSS J1553+0056 show relative to
LBGs (1) a significantly different flux distribution in the (rest-frame)
wavelength range 1550 -- 1800 \AA, (2) stronger low-ionization resonance
absorption features, and (3) stronger N\,{\sc v} resonance absorption. If any
of these properties is observed in a low-resolution LBG spectrum, the object
could be a misidentified BALQSO. However, because of the large range of
properties of BALQSOs, the absence of any of these signs does not guarantee
that the observed object is a bona fide LBG. Particularly at the high
luminosity end of the LBG luminosity function, there is obviously the danger
of BALQSOs polluting low-resolution spectroscopy based LBG samples.
 
\begin{acknowledgements}
It is a pleasure to thank the ESO Paranal Observatory staff for
carrying out for us the service mode observations on which this
paper is based, and Dr. Patrick Osmer for reading our manuscript and
for valuable comments. This work was supported by the German Science 
Foundation (DFG, SFB 439).
\end{acknowledgements}

\bibliographystyle{aa}
\bibliography{qso}

\end{document}